\documentclass{ws-ijmpe}
\usepackage[super,compress]{cite}

\begin{document}

\markboth{M. Thoennessen}{2013 Update of the Discoveries of Isotopes}

\catchline{}{}{}{}{}

\title{2013 UPDATE OF THE DISCOVERIES OF NUCLIDES}

\author{\footnotesize M. THOENNESSEN}

\address{National Superconducting Cyclotron Laboratory and \\
Department of Physics \& Astronomy \\
Michigan State University\\
East Lansing, Michigan 48824, USA\\
thoennessen@nscl.msu.edu}

\maketitle

\begin{history}
\received{Day Month Year}
\revised{Day Month Year}
\end{history}

\begin{abstract}
The 2013 update of the discovery of nuclide project is presented. Details of the 12 new nuclides observed for the first time in 2013 are described. In addition, the discovery of $^{266}$Db has been included and the previous assignments of 6 other nuclides were changed. Overview tables of where and how nuclides were discovered have also been updated and are discussed.
\end{abstract}


\ccode{PACS numbers: 21.10.-k, 29.87.+g}


\section{Introduction}

The initial idea for the project to document the discovery of all nuclides originated from a review article discussing nuclides at the limit of stability\cite{2004Tho01}. In this article the discovery of nuclides near and beyond the proton (12 $\le$ Z $\le$ 97) and neutron (9 $\le$ N $\le$ 36) dripline were presented. Later the range was expanded to include the lighter nuclides (4 $\le$ Z and 4 $\le $ N)\cite{2005Tho01,2005Tho02}. The more systematic approach to describe and discuss the discovery of all nuclides began in 2007 and the first paper on the discovery of all cerium isotopes was submitted to At. Data  Nucl. Data Tables in 2008.\cite{2009Gin01} Papers on the isotopes of arsenic,\cite{2010Sho03} gold,\cite{2010Sch02} tungsten,\cite{2010Fri01} and  krypton,\cite{2010Hei01} followed in 2009 until the project was completed in November 2011 with the submission of the final paper on the discovery of the actinium, thorium, protactinium, and uranium isotopes.\cite{2013Fry04} A complete list of the discovery of all isotopes of all elements can be found in Ref. \refcite{2013Tho02} and on the web.\cite{2011Tho03}

The criteria for the assignments of a discovery are certainly debatable. For the current project the following main criteria were used: (1) clear identification, either through decay curves and relationships to other known nuclides, particle or $\gamma$-ray spectra, or unique mass and element identification, and (2) published in a refereed journal.\cite{2013Tho02} Also, the definition of what constitutes a nucleus is not well defined. For the discovery project all nuclides which exist for more than $\sim$10$^{-22}$s which can be considered a characteristic nuclear timescale\cite{2004Tho01,1960Gol01} were included. As in the previous review the correct word ``nuclide'' is used rather than the colloquial term ``isotope'' which strictly speaking should only be used for nuclides of a specific element.\cite{2013Tho02}

\section{New discoveries in 2013}

In 2013, the discoveries of only 12 new isotopes were reported in refereed journals. They are listed in Table \ref{2013Isotopes}. One neutron unbound resonance, 6 neutron-rich nuclei close to the neutron dripline, two proton-rich $\alpha$-emitting nuclei and two transuranium nuclides at the end of super-heavy nuclei decay chains. 

Although $^{15}$Be was expected to be unbound with respect to neutron emission because the heavier isotone $^{16}$B had been shown to be unbound \cite{1973Bow01}, it was experimentally confirmed only recently.\cite{2011Spy01} Snyder {\it et al.} measured an unbound resonance in $^{15}$Be from the invariant mass reconstruction of $^{14}$Be fragments and neutrons measured in coincidence: ``It was populated with neutron transfer from a deuterated polyethylene target in inverse kinematics with a radioactive beam of $^{14}$Be. $^{15}$Be decays by neutron emission to $^{14}$Be with a decay energy of 1.8(1)~MeV.''\cite{2013Sny01} 

The neutron-rich nuclides $^{64}$Ti, $^{67}$V, $^{69,70}$Cr, $^{72}$Mn, $^{75}$Fe were identified for the first time in a projectile fragmentation reaction from a 139 MeV/nucleon primary $^{82}$Se beam: ``The observed fragments include several new isotopes that are the most neutron-rich nuclides yet observed of elements 22 $\leq$ Z $\leq$ 25 ($^{64}$Ti,$^{67}$V, $^{69}$Cr, and $^{72}$Mn). One event was found to be consistent with $^{70}$Cr, and another one was found to be consistent with $^{75}$Fe.''\cite{2013Tar01}

$^{131}$Ag is the first new nuclide that was produced from a fragmentation reaction from a secondary beam. A 345 MeV/nucleon primary $^{235}$U beam was first fragmented to produce an intense secondary $\sim$230 MeV/nucleon $^{134,135}$Sn  beam which then was fragmented again to produce neutron-rich silver isotopes: ``In total, 30 events for $^{131}$Ag were identified by measuring its magnetic rigidity, time of flight, energy loss and total kinetic energy.''\cite{2013Wan01}

Kalaninova {\it et al.} reported the discovery of $^{197}$Fr produced in the fusion-evaporation reaction $^{141}$Pr($^{60}$Ni,4n): ``The new isotope $^{197}$Fr was identified based on the observation of one $\alpha$-decay chain yielding E$_\alpha$ = 7728(15)~keV and T$_{1/2}$ = 0.6$^{+3.0}_{-0.3}$~ms.''\cite{2013Kal01} They also observed the decay of $^{198}$Fr but did not claim the discovery stating: ``Recently, an $\alpha$-decay study of $^{198,199}$Fr was performed at the gas-filled separator Recoil Ion Transport Unit (RITU) at The University of Jyv\"askyl\"a (JYFL) [J. Uusitalo (private communication)]. However, to our knowledge, no results were published so far.''\cite{2013Kal01} they did not claim the discovery of $^{198}$Fr. 

Apparently unbeknownst to Kalaninova {\it et al.} Uusitalo {\it et al.} had submitted their results for the discovery of $^{198}$Fr (Ref.~\refcite{2013Uus01}) two months earlier. Residues produced in the fusion-evaporation reaction $^{141}$Pr($^{60}$Ni,3n) were identified with the gas-filled recoil separator RITU: ``Two $\alpha$-particle activities, with E$_\alpha$ = 7613(15)~keV and T$_{1/2}$ = (15$^{+12}_{-5}$)~ms and E$_\alpha$ = 7684(15)~keV and T$_{1/2}$ = (16$^{+13}_{-5}$)~ms were identified in the new isotope $^{198}$Fr.''\cite{2013Uus01}

The group of Oganessian {\it et al.} continued their studies of producing superheavy nuclei in hot fusion-evaporation reactions and reported the first observation of $^{271}$Bh and $^{277}$Mt  at the end of the decay chains originating from $^{287}$115 and $^{293}$117, respectively. One decay chain was observed in the reaction $^{243}$Am($^{48}$Ca,4n)$^{287}$115: ``Here, the $\alpha$-decay energy and lifetime of $^{271}$Bh were detected for the first time.''\cite{2013Oga01} 
The reaction $^{249}$Bk($^{48}$Ca,4n) was used to study the decay chain of $^{293}$117: ``An $\alpha$-decay branch of $^{281}$Rg leading to the new SF nucleus, $^{277}$Mt, was observed for the first time.''\cite{2013Oga02}

\begin{table}[pt]
\tbl{New nuclides reported in 2013. The nuclides are listed with the first author, submission date, and reference of the publication, the laboratory where the experiment was performed, and the production method (SB = secondary beams, PF = projectile fragmentation, FE = fusion evaporation). \label{2013Isotopes}}
{\begin{tabular}{@{}llcclc@{}} \toprule 
Nuclide(s) & Author & Subm. Date & Ref. & Laboratory & Type \\ \colrule
$^{15}$Be&  J. Snyder et al.  & 4/22/2013 &  \refcite{2013Sny01} &  MSU   & SB  \\
\parbox[t]{2.5cm}{\raggedright $^{64}$Ti, $^{67}$V, $^{69,70}$Cr, $^{72}$Mn, $^{75}$Fe \vspace*{0.1cm}}&  O. Tarasov et al.  & 2/11/2013 &  \refcite{2013Tar01} &  MSU  & PF  \\
$^{131}$Ag&  H. Wang et al. & 12/11/2012 &  \refcite{2013Wan01}  &  RIKEN   & SB \\
$^{197}$Fr&  Z. Kalaninova et al.  & 1/29/2013 &  \refcite{2013Kal01} &  GSI  & FE \\
$^{198}$Fr&  J. Uusitalo et al.   & 11/30/2012 &  \refcite{2013Uus01} &  Jyv\"askyl\"a   & FE \\
$^{271}$Bh&  Yu. Ts. Oganessian et al.  & 9/1/2012 &  \refcite{2013Oga01} &  Dubna    & FE \\
$^{277}$Mt&  Yu. Ts. Oganessian et al.  & 4/6/2013 &  \refcite{2013Oga02} &  Dubna    & FE \\
\botrule
\end{tabular}}
\end{table}

\section{Changes of prior assignments}

In addition to the 12 new nuclides discovered in 2013 some of the previous assignments were reevaluated. In this process, the discovery of $^{266}$Db was accepted. Originally it had been rejected: ``$^{266}$Db was at the end of the isotope chain originating at $^{282}$113, however, the observed spontaneous fission could have been due to either $^{266}$Db or $^{266}$Rf [Ref. \refcite{2007Oga01}] and [Ref. \refcite{2007Oga02}].''\cite{2013Tho01} In a recent paper Oganessian {\it et al.} indicated that the spontaneous fission is indeed occurring from $^{266}$Db. In Figure 5 of Ref. \refcite{2013Oga01} they show $^{266}$Db as decaying by fission with a half-life of 22$^{+105}_{-10}$~min and state in the caption: ``For five new spontaneously fissioning nuclei marked by gray squares, that is, the isotopes of Db and Rg that terminate the $\alpha$-decay sequences with SF, the half-lives are listed.''\cite{2013Oga01} Another reason to accept the discovery is that the measured half-life was assigned to $^{266}$Db independent of its decay mode: ``The first decay chain was terminated by SF decay with an apparent life time of 31.7~min. The origin of this decay can be spontaneous fission of $^{266}$Db, or its $\epsilon$ decay with a life time of 31.7~min followed by the relatively short-lived spontaneous fission of the even-even isotope $^{266}$Rf.''\cite{2007Oga01}

In addition to the acceptance of the $^{266}$Db a few of the previous discoveries were reassigned. In the early 1970s several papers identified nuclides by assigning $\gamma$-rays to fission fragments detected in spontaneous fission of $^{252}$Cf.\cite{1970Che01,1970Wat01,1970Wil01,1971Hop01,1972Hop01} In the course of analyzing the different elements these claims were not treated uniformly. A consistent review of these papers led to the following three reassignments: $^{101}$Zr was reassigned from Trautmann {\it et al.}\cite{1972Tra01} to Watson {\it et al.}\cite{1970Wat01}, $^{108}$Tc from Watson {\it et al.}\cite{1970Wat01} to Trautmann {\it et al.}\cite{1972Tra01} and $^{149}$Ce from Aronsson {\it et al.}\cite{1974Aro01} to Hopkins {\it et al.}\cite{1971Hop01}.

The discovery of $^{195}$Os had previously been assigned to Valiente-Dobon {\it et al.}\cite{2004Val01} because an earlier observation by Bar\'o and Rey\cite{1957Bar01} had been rejected because of a statement in an annual report: ``Unfortunately, the then-existing assignment for $^{195}$Ir has subsequently been identified as $^{81}$Rb, arising from reactions induced in target impurities. As a result, the present assignment of $^{195}$Os will not withstand careful scrutiny.''\cite{1974Col01} However, Birch {\it et al.} recently pointed out that this rejection was not justified and  Bar\'o and Rey should be credited with the discovery of $^{195}$Os.\cite{2013Bir01}

In the paper describing the discovery of actinium isotopes\cite{2013Fry04} the discovery of $^{235}$Ac had been assigned to  Bosch {\it et al.}\cite{2006Bos01}. However, Ta\"ieb {\it et al.} had reported the first observation of $^{235}$Ac three years earlier: ``We observed, for the first time, the isotope $^{235}$Ac that corresponds to the 3-proton removal channel. 150 events were unambiguously recorded.''\cite{2003Tai01} Unfortunately, this paper had been overlooked and so the assignment has been changed to credit Ta\"ieb {\it et al.} for the discovery of  $^{235}$Ac.
 
Finally, the last case is $^{271}$Ds which has been really difficult to assign. The first time decay properties of $^{271}$Ds were reported in the refereed literature was a single-author review article by Hofman\cite{1998Hof01}. The fact that it had been observed had been mentioned in a separate paper earlier: ``In a succeeding experiment we investigated the reaction $^{64}$Ni + $^{208}$Pb and observed the heavier isotope $^{271}$110,''\cite{1995Hof02} however no details were given referencing a non-refereed publication.\cite{1994Hof01} The decay chain actually had been published in a refereed publication in 1995\cite{1995Arm01} although again not all researchers who participated in the experiment were co-authors of the paper. At the current time Ref.~\refcite{1995Arm01} is credited with the discovery of $^{271}$Ds although it probably would be more appropriate to make an exception to the ``refereed publication'' rule and assign the discovery to the announcement in the ``GSI Nachrichten''\cite{1994Hof01} in order to give credit to all researchers involved in the discovery.

\section{Status at the end of 2013}

\begin{figure}[pt] 
\centerline{\psfig{file=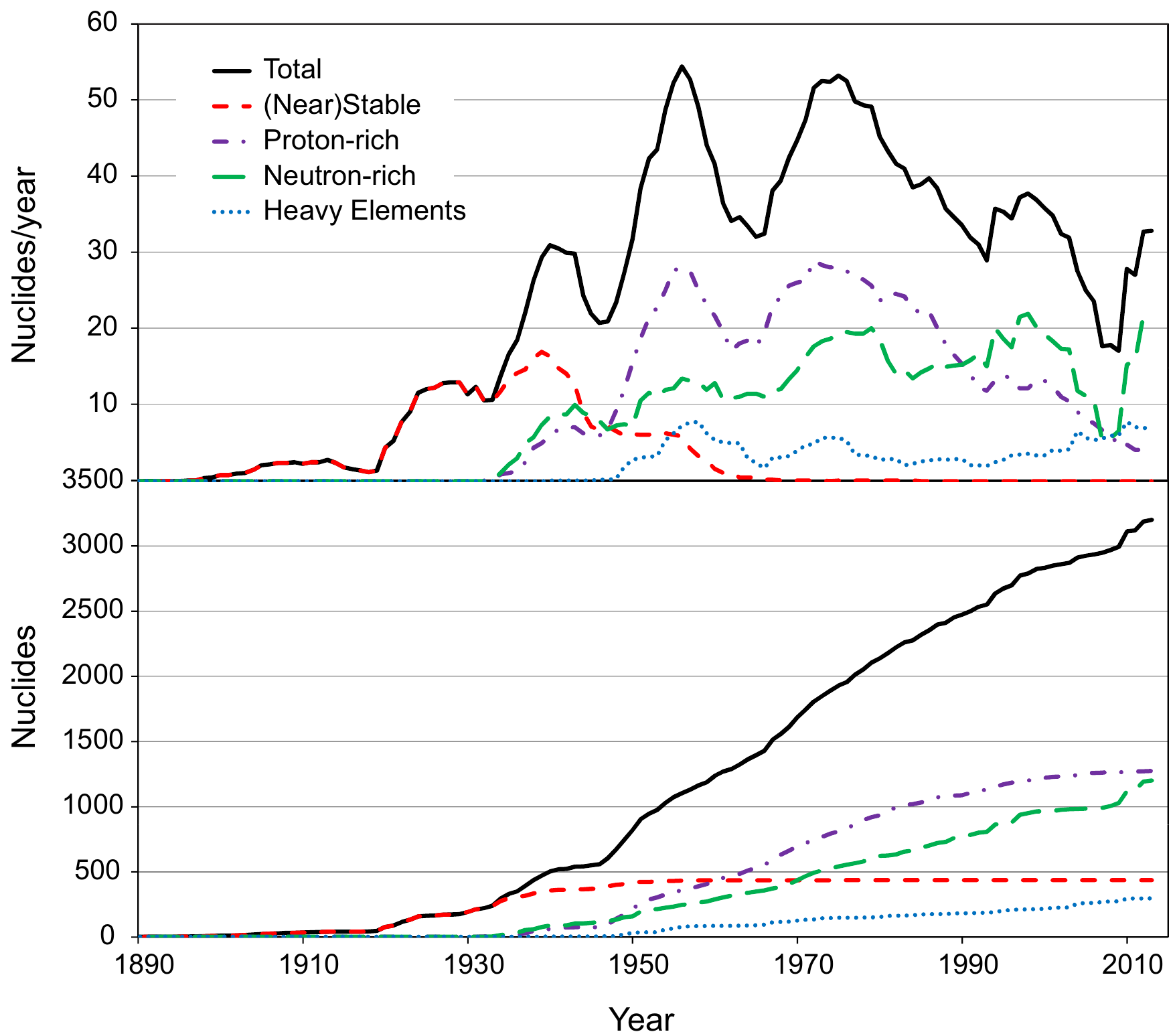,width=12cm}}
\caption{Discovery of nuclides as a function of year. The top figure shows the 10-year running average of the number of nuclides discovered per year while the bottom figure shows the cumulative number.  The total number of nuclides shown by the black, solid lines are plotted separately for near-stable (red, short-dashed lines), neutron-deficient (purple, dot-dashed lines), neutron-rich (green, long-dashed lines) and transuranium (blue, dotted lines) nuclides. This figure was adapted from Ref. \cite{2013Tho02} to include the data from the two most recent years (2012 and 2013). \label{f:timeline}}
\end{figure}

With the new discoveries and the few reassignments the current status of the evolution of the nuclide discovery is shown in Figure \ref{f:timeline}. The figure was adapted from the previous review\cite{2013Tho02} and extended to include the two most recent years of 2012 and 2013. The top part of the figure shows the ten-year average of the number of nuclides discovered per year while the bottom part of the figure shows the integral number of nuclides discovered. It can be seen that the recent rate increase that started in 2010 continued and that the current 2013 rate of 32.9 nuclides/year is the largest number since 2001. This increase is primarily due to the large number of new neutron-rich nuclides being discovered at fragmentation facilities. The current rate of 22.0 neutron-rich nuclides discovered per year equals the largest rate first reached in 1998.

The rate of new proton-rich nuclei continues to decline while the rate of the discovery of nuclides of heavy elements is fairly constant at a high level of 7.0. The highest rate of 7.8 was set in the late-fifties and had been equaled in 2010.

Overall 3195 different nuclides have been discovered so far. While the number of proton-rich nuclides is saturating close to 1300 nuclides (1271) the number of neutron-rich nuclides continues to increase (1197) and will most likely surpass the number of proton-rich nuclides within the next few years. These numbers include 19 neutron- and about 40 proton-unbound nuclides. While this distinction is straightforward along the neutron dripline, it is less clear along the proton dripline due to presence of the Coulomb barrier. 

From Figure \ref{f:timeline} it is obvious that the discovery rate has not proceeded steadily at a constant rate but that it exhibits large fluctuations. It already had been pointed out that periods of high discovery rates are correlated with the development of new methods, instrumentation, techniques or accelerators.\cite{2011Tho01,2011Rei01,2011TrW01,2011Tho02}

\begin{table}[pt] 
\tbl{Top ten countries where the most nuclides were discovered. The total number of nuclides are listed together with first and most recent year of a discovery. \label{countries}}
{\begin{tabular}{@{}rlrcc@{}} \toprule
Rank & Country & Number & First year & Recent year \\ \colrule
1	&	 USA 	&	1324	&	1907	&	2013	\\
2	&	 Germany 	&	554	&	1898	&	2013	\\
3	&	 UK 	&	300	&	1900	&	1994	\\
4	&	 Russia 	&	248	&	1957	&	2013	\\
5	&	 France 	&	213	&	1896	&	2005	\\
6	&	 Switzerland 	&	129	&	1934	&	2009	\\
7	&	 Japan 	&	127	&	1938	&	2013	\\
8	&	 Sweden 	&	61	&	1945	&	1993	\\
	&	 Canada 	&	61	&	1900	&	1998	\\
10	&	 Finland 	&	38	&	1961	&	2013	\\
\botrule
\end{tabular}}
\end{table}

Overview tables of the demographics of the nuclide discoveries have been published for 2011\cite{2012Tho03} and yearly updates are available on the web.\cite{2011Tho03}

The 3195 nuclides were discovered in 25 different countries. Table \ref{countries} lists the 10 countries where the most nuclides were discovered. There were essentially no changes over the last two years.\cite{2011Tho03,2012Tho03} The reassignment of the discovery of $^{149}$Ce from Aronsson {\it et al.}\cite{1974Aro01} to Hopkins {\it et al.}\cite{1971Hop01} reduced the number for Sweden from 62 to 61 moving Canada from ninth to a tie with Sweden at eighth. By far the most were discovered at laboratories within the USA. This includes $^{201}$Pt which was first observed at the nuclear reactor of the Puerto Rico Nuclear Center in Mayag\"uez in Puerto Rico.\cite{1962Fac01} The number for Germany includes nuclides discovered in West-Germany between 1949 and 1990. Also, the number for Germany and France should be increased by 0.5 because $^{105}$Nb was reported in a paper describing results from the fission product separators JOSEF and LOHENGRIN at J\"ulich and Grenoble, respectively.\cite{1984Shi01} The total number of nuclides in Russia include discoveries from 1957 through 1991 in the USSR including $^{94}$Rh (Ref.~\refcite{1979Zar01}) and $^{106}$Sn (Ref.~\refcite{1975Bur01}) which were first synthesized at the isochronous cyclotron of the Nuclear Physics Institute of the Kazakh Academy of Sciences in Almaty, Kaszakhstan.  It is interesting to note that with the exception of the UK, Sweden and Canada, all countries reported new nuclides within the last 8 years.

Over 120 different laboratories measured new nuclides. The top ten laboratories are listed in Table \ref{labs}. The most nuclides were discovered at Berkeley starting with $^{21}$Ne identified in 1928 in the Chemical Laboratories at the University of California,\cite{1928Hog01} and $^{15}$O produced in 1934 in the Radiation Laboratory.\cite{1934Liv01} The two most recent discoveries of $^{271}$Bh (Ref.~\refcite{2013Oga01}) and $^{277}$Mt (Ref.~\refcite{2013Oga02}) moved Dubna ahead of Cambridge. Also, with the 7 new nuclides discovered in 2013\cite{2013Sny01,2013Tar01} Michigan State moved into the top ten tied with RIKEN. Similar to the countries, with the exception of Cambridge and Orsay, all leading laboratories reported new nuclides within the last 8 years.

\begin{table}[pt] 
\tbl{Top ten laboratories where the most nuclides were discovered. The total number of nuclides are listed together with the country of the laboratory and the first and most recent year of a discovery. \label{labs}}
{\begin{tabular}{@{}rllrcc@{}} \toprule
Rank & Laboratory & Country & Number & First year & Recent year \\ \colrule
1	&	 Berkeley 	& USA		&	635	&	1928	&	2010	\\
2	&	 GSI	 	& Germany	&	435	&	1977	&	2013	\\
3	&	 Dubna 	& Russia	&	223	&	1957	&	2013	\\
4	&	 Cambridge 	& UK		&	222	&	1913	&	1940	\\
5	&	 CERN 	& Switzerland&	119	&	1965	&	2009	\\
6	&	 Argonne 	& USA		&	118	&	1947	&	2012	\\
7	&	 GANIL 	& France	&	85	&	1985	&	2005	\\
8	&	 Oak Ridge 	& USA		&	77	&	1946	&	2006	\\
9	&	 Orsay 	& France	&	73	&	1959	&	1989	\\
10	&	 Michigan State 	& USA &	72	&	1967	&	2013	\\
	&	 RIKEN 	& Japan	&	72	&	1972	&	2013	\\
\botrule
\end{tabular}}
\end{table}

The experiments performed to produce and identify new nuclides involved more than 3400 researchers who published their results in 1521 different publications. Table \ref{coauthor} lists the top ten reseachers who are coauthors on publications of the most discoveries. Last year, as coauthors of a paper by Kurcewicz {\it et al.}\cite{2012Kur01} in which 59 new nuclides were reported Geissel and Pf\"utzner took over the top two spots.\cite{2013GSI01} Previously, for the last eight years M\"unzenberg was listed as coauthor on the most discovery papers. This year there were only minor changes in the top ten with T. Kubo from RIKEN and D. Bazin from Michigan State (MSU) moving to ninth and S. Czajkowski (GSI/Bordeaux) and C. Donzaud (Orsay/GSI) dropping out of the top ten. The list is dominated by researchers involved in projectile fragmentation experiments which are able to identify a large number of nuclides within one setting of the fragment separator and which typically involve many researchers. Thus it is no surprise that with the exception of Aston from Cambridge and Bazin from MSU all researchers participated in nuclide discoveries at GSI.

\begin{table}[pt] 
\tbl{Top ten researchers listed as coauthors of publications of nuclide discoveries. The total number of nuclides are listed together with the laboratories where the experiments were performed. \label{coauthor}}
{\begin{tabular}{@{}rllr@{}} \toprule
Rank & Country & Labs & Number \\ \colrule
1	&	 H. Geissel 	& GSI/RIKEN	& 271	\\
2	&	 M. Pf\"utzner 	& GSI/GANIL	&	224	\\
3	&	 G. M\"unzenberg 	& GSI	&	218	\\
4	&	 F. W. Aston 	& Cambridge	& 207	\\
5	&	 P. Armbruster 	& GSI/Grenoble/J\"ulich/Jyv\"askyl\"a &	203	\\
6	&	 M. Bernas 	& GSI/GANIL/Grenoble/Orsay	&	164	\\
7	&	 K. S\"ummerer 	& GSI/GANIL	&	153	\\
8	&	 A. Heinz 	& GSI/GANIL/ANL	&	147	\\
9	&	 T. Kubo 	& GSI/RIKEN/MSU	&	137	\\
	&	 D. Bazin 	& GANIL/RIKEN/MSU	&	137	\\
\botrule
\end{tabular}}
\end{table}

The situation is different if one considers only first-author publications. Table \ref{first} lists the top ten researchers who are first authors on publications of the most discoveries. Although there are still five researchers from projectile fragmentation experiments on the list (Bernas, Kurcewicz, Ohnishi, Alvarez-Pol, and Guillemaud-Mueller) Aston who pioneered mass spectroscopy of stable nuclei leads by a large margin. It demonstrates the fact that during the early parts of nuclear physics, the experiments were performed by a single researcher.\cite{2012Tho03}

\begin{table}[pt] 
\tbl{Top ten researchers listed as first authors of publications of nuclide discoveries. The total number of nuclides are listed together with the laboratories where the experiments were performed. \label{first}}
{\begin{tabular}{@{}rllr@{}} \toprule
Rank & Country & Labs & Number \\ \colrule
1	&	 F. W. Aston 	& Cambridge	&	207	\\
2	&	 M. Bernas 	& GSI	&	110	\\
3	&	 J. Kurcewicz & GSI	&	59	\\
4	&	 Yu. Ts. Oganessian & Dubna	&	50	\\
5	&	 T. Ohnishi 	& RIKEN	&	49	\\
6	&	 H. Alvarez-Pol & GSI	&	39	\\
7	&	 K. S. Toth 	& ORNL/LBL/TAMU/ANL/Uppsala	&	37	\\
8	&	 S. Hofmann 	 & GSI	&	36	\\
	&	 A. J. Dempster 	& Chicago &	36	\\
10	&	 D. Guillemaud-Mueller 	& GANIL	&	35	\\
\botrule
\end{tabular}}
\end{table}

\begin{table}[pt] 
\tbl{Top ten experimental methods to produce and identify new nuclides. The total number of nuclides are listed together with the first and most recent year of a discovery. \label{method}}
{\begin{tabular}{@{}rlrcc@{}} \toprule
Rank & Country & Number & First year & Recent year \\ \colrule
1	&	 Light particle induced 	&	769	&	1925	&	2003	\\
2	&	 Fusion evaporation 	&	746	&	1951	&	2013	\\
3	&	 Projectile fission/fragmentation 	&	493	&	1979	&	2013	\\
4	&	 Mass spectroscopy 	&	270	&	1908	&	1949	\\
5	&	 Spallation/ target fragmentation 	&	247	&	1948	&	2009	\\
6	&	 Neutron induced fission 	&	241	&	1939	&	1993	\\
7	&	 Neutron capture 	&	125	&	1934	&	1987	\\
8	&	 Deep inelastic 	&	65	&	1970	&	2010	\\
9	&	 Charged particle fission 	&	51	&	1948	&	2008	\\
	&	 Radioactive decay 	&	51	&	1896	&	1961	\\
\botrule
\end{tabular}}
\end{table}

Another way to categorize the discovery experiments is to sort them by production mechanism or experimental technique.  Table \ref{method} lists the top ten different methods used to populate and identify the nuclides for the first time. Nuclides that are still unknown are further and further removed from the valley of stability and only a few methods remain viable to reach these nuclei and study their properties. In the last five years only five different methods have succeeded in producing new nuclides. In addition to fusion evaporation, projectile fission/fragmentation, spallation/target fragmentation and deep inelastic reactions which are listed in the top ten, varies reactions with secondary beams were able to reach new nuclei. The utilization of reactions induced by secondary beams is the newest - and arguably the most difficult - method which was first applied in 1994 in the discovery of $^{10}$He at RIKEN by Korsheninnikov {\it et al.}\cite{1994Kor01} and produced ten new nuclides during the last five years.

The last nuclide discovered by mass spectroscopy was $^{50}$V which was simultaneously reported  by Hess and Inghram\cite{1949Hes01} from Argonne National Laboratory and the University of Chicago and Leland\cite{1949Lel01} from the University of Minnesota. It is interesting to note that the last naturally occurring radioactive nucleus was only discovered in 1961 by Nurmia {\it et al.} in Helsinki, Finland.\cite{1961Nur01} They measured the decay of $^{206}$Bi following the $\alpha$-decay of $^{210}$Pb with a branching ratio of $\sim$10$^{-9}$.

Finally, from a historical viewpoint it is interesting to see in which journals most of the discoveries were reported. Table \ref{journal} lists the top ten different journals which published the most discovery papers. By far the most discoveries (over 1000) were reported in Physical Review and Physical Review C (at the end of 1969 Physical Review split up into several different journals and from 1970 Nuclear Physics was continued in Physical Review C). In addition, starting in 1958 another $\sim$200 nuclides were first reported in Physical Review Letters, another journal of the American Physical Society. In Europe, in the early days of Nuclear Physics many of the new discoveries were reported in Nature. However, after 1961 Nature did not publish any discovery papers until the 2007 publication of the first observation of $^{40}$Mg and $^{42,43}$Al (Ref. \refcite{2007Bau01})  demonstrating the renewed broader interest in Nuclear Physics. Within the classified research of the Manhattan Project during WWII many new nuclides were discovered. The papers describing these discoveries were later unclassified and published in several volumes of the National Nuclear Energy Series (Nat. Nucl. Ener. Ser.).

\begin{table}[pt]
\tbl{Top ten journals where new nuclides were reported. The total number of nuclides are listed together with the first and most recent year of a discovery. \label{journal}}
{\begin{tabular}{@{}rlrcc@{}} \toprule
Rank & Country & Number & First year & Recent year \\ \colrule
1	&	 Phys. Rev. 	&	737	&	1922	&	1969	\\
2	&	 Phys. Rev. C 	&	366	&	1970	&	2013	\\
3	&	 Phys. Lett. B 	&	322	&	1967	&	2012	\\
4	&	 Z. Phys. A 	&	304	&	1975	&	1997	\\
5	&	 Nature 	&	260	&	1905	&	2012	\\
6	&	 Nucl. Phys. A 	&	225	&	1967	&	2003	\\
7	&	 Phys. Rev. Lett. 	&	211	&	1958	&	2012	\\
8	&	 J. Inorg. Nucl. Chem. 	&	93	&	1955	&	1981	\\
9	&	 J. Phys. Soc. Japan 	&	62	&	1960	&	2010	\\
10	&	 Nat. Nucl. Ener. Ser. 	&	49	&	1949	&	1951	\\
\botrule
\end{tabular}}
\end{table}

\section{Discoveries not yet published in refereed journals}

\begin{table}[pt]
\tbl{Nuclides only reported in proceedings or internal reports until the end of 2013. The nuclide, first author, year, laboratory, conference or report and reference of the discovery are listed. \label{reports}}
{\begin{tabular}{@{}lllllr@{}} \toprule
\parbox[t]{1.5cm}{\raggedright Nuclide(s) \vspace*{0.2cm}}& Author & Year & Lab. & Conf./Report & Ref.\\ \colrule
\parbox[t]{1.5cm}{\raggedright $^{81,82}$Mo, $^{85,86}$Ru \vspace*{0.2cm}}&  H. Suzuki et al.  & 2013 &  RIKEN  &  \parbox[t]{2.8cm}{\raggedright EMIS 2012 \vspace*{0.2cm}}  &  \refcite{2013Suz01} \\$^{156}$Pr&  S. Czajkowski et al.  & 1996 &  GSI  &  \parbox[t]{2.8cm}{\raggedright ENAM'95  \vspace*{0.2cm}}  &  \refcite{1996Cza01} \\
\parbox[t]{1.5cm}{\raggedright $^{126}$Nd, $^{136}$Gd, $^{138}$Tb  \vspace*{0.2cm}}&  G. A. Souliotis  & 2000 &  MSU  &  \parbox[t]{2.8cm}{\raggedright Achiev. and Persp. in Nucl. Struct. 1999 \vspace*{0.2cm}}  &  \refcite{2000Sou01} \\
\parbox[t]{1.5cm}{\raggedright$^{143}$Ho \vspace*{0.2cm}}  &  G. A. Souliotis  & 2000 &  MSU  &  \parbox[t]{2.8cm}{\raggedright Achiev. and Persp. in Nucl. Struct. 1999 \vspace*{0.08cm}}  &  \refcite{2000Sou01} \\
  &  D. Seweryniak et al.  & 2002 &  LBL  &  \parbox[t]{2.8cm}{\raggedright Annual Report \vspace*{0.2cm}}  &  \refcite{2003Sew02} \\
\parbox[t]{1.5cm}{\raggedright$^{144}$Tm \vspace*{0.2cm}}  &  K. P. Rykaczewski et al.  & 2004 &  ORNL  &  \parbox[t]{2.8cm}{\raggedright Nuclei at the Limits 2004 \vspace*{0.08cm} }  &  \refcite{2005Ryk01} \\
 &  R. Grzywacz et al.  &   &   &  \parbox[t]{2.8cm}{\raggedright ENAM2004 \vspace*{0.05cm}}  &  \refcite{2005Grz01} \\
 &  C. R. Bingham et al.  &   &   &  \parbox[t]{2.8cm}{\raggedright CAARI2004 \vspace*{0.2cm}}  &  \refcite{2005Bin01} \\
\parbox[t]{1.5cm}{\raggedright $^{150}$Yb, $^{153}$Hf   \vspace*{0.2cm}} &  G. A. Souliotis  & 2000 &  MSU  &  \parbox[t]{2.8cm}{\raggedright Achiev. and Persp. in Nucl. Struct. 1999 \vspace*{0.2cm}}  &  \refcite{2000Sou01} \\
\parbox[t]{1.5cm}{\raggedright$^{164}$Ir \vspace*{0.2cm}}  &  H. Kettunen et al.  & 2000 &  Jyv\"askyl\"a  &  \parbox[t]{2.8cm}{\raggedright Zakopane School of Physics 2000 \vspace*{0.08cm}}  &  \refcite{2001Ket02} \\
 &  H. Mahmud et al.  & 2001 &  ANL  &  \parbox[t]{2.8cm}{\raggedright ENAM2001 \vspace*{0.05cm} }  &  \refcite{2002Mah01} \\
 &  D. Seweryniak et al.  &   &   &  \parbox[t]{2.8cm}{\raggedright Frontiers of Nuclear Structure 2002 \vspace*{0.2cm}}  &  \refcite{2003Sew01} \\
\parbox[t]{1.5cm}{\raggedright $^{230}$At, $^{232}$Rn   \vspace*{0.2cm}} &  J. Benlliure et al.  & 2010 &  GSI  &  \parbox[t]{2.8cm}{\raggedright arXiv \\ NIC XI \vspace*{0.2cm}}  &  \parbox[t]{1cm}{\raggedright\refcite{2010Ben01} \refcite{2010Ben02}} \\
\parbox[t]{1.5cm}{\raggedright$^{234}$Cm \vspace*{0.08cm}}  &  P. Cardaja et al.  & 2002 &  GSI  &  \parbox[t]{2.8cm}{\raggedright Annual Report \vspace*{0.05cm}}  &  \refcite{2002Cag01} \\
 &  J. Khuyagbaatar et al.  & 2007 &  GSI  &  \parbox[t]{2.8cm}{\raggedright Annual Report \vspace*{0.05cm}}  &  \refcite{2007Khu01} \\
 &  D. Kaji et al.  & 2010 &  RIKEN  &  \parbox[t]{2.8cm}{\raggedright Annual Report \vspace*{0.2cm}}  &  \refcite{2010Kaj01} \\
\parbox[t]{1.5cm}{\raggedright$^{235}$Cm \vspace*{0.2cm}}  &  J. Khuyagbaatar et al.  & 2007 &  GSI  &  \parbox[t]{2.8cm}{\raggedright Annual Report \vspace*{0.2cm}}  &  \refcite{2007Khu01} \\
\parbox[t]{1.5cm}{\raggedright$^{234}$Bk \vspace*{0.2cm}}  &  K. Morita et al.  & 2002 &  RIKEN  &  \parbox[t]{2.8cm}{\raggedright Front. of Coll. Motion 2002 \vspace*{0.1cm}}  &  \refcite{2003Mor02} \\
 &  K. Morimoto et al.  &   &   &  \parbox[t]{2.8cm}{\raggedright Annual Report \vspace*{0.05cm}}  &  \refcite{2003Mor01} \\
 &  D. Kaji et al.  & 2010 &  RIKEN  &  \parbox[t]{2.8cm}{\raggedright Annual Report \vspace*{0.2cm}}  &  \refcite{2010Kaj01} \\
\parbox[t]{1.5cm}{\raggedright$^{252,253}$Bk \vspace*{0.2cm}}  &  S. A. Kreek et al.  & 1992 &  LBL  &  \parbox[t]{2.8cm}{\raggedright Annual Report \vspace*{0.2cm} }  &  \refcite{1992Kre01} \\
\parbox[t]{1.5cm}{\raggedright$^{262}$No \vspace*{0.08cm}}  &  R. W. Lougheed et al.  & 1988 &  LBL  &  \parbox[t]{2.8cm}{\raggedright Annual Report }  &  \refcite{1988Lou01} \\
 &   &   &   &  \parbox[t]{2.8cm}{\raggedright 50 years with nuclear fission 1989 \vspace*{0.08cm}}  &  \refcite{1989Lou01} \\
 &  E. K. Hulet  &   &   &  \parbox[t]{2.8cm}{\raggedright Internal Report \vspace*{0.2cm} }  &  \refcite{1989Hul01} \\
\parbox[t]{1.5cm}{\raggedright$^{261}$Lr  \vspace*{0.08cm}} &  R. W. Lougheed et al.  & 1987 &  LBL  &  \parbox[t]{2.8cm}{\raggedright Annual Report \vspace*{0.05cm}}  &  \refcite{1987Lou01} \\
 &  E. K. Hulet  &   &   &  \parbox[t]{2.8cm}{\raggedright Internal Report \vspace*{0.05cm}}  &  \refcite{1989Hul01} \\
 &  R. A. Henderson et al.  & 1991 &  LBL  &  \parbox[t]{2.8cm}{\raggedright Annual Report \vspace*{0.2cm}}  &  \refcite{1991Hen01} \\
\parbox[t]{1.5cm}{\raggedright$^{262}$Lr \vspace*{0.08cm}}  &  R. W. Lougheed et al.  & 1987 &  LBL  &  \parbox[t]{2.8cm}{\raggedright Annual Report \vspace*{0.05cm}}  &  \refcite{1987Lou01} \\
 &  E. K. Hulet  &   &   &  \parbox[t]{2.8cm}{\raggedright Internal Report \vspace*{0.05cm}}  &  \refcite{1989Hul01} \\
 &  R. A. Henderson et al.  & 1991 &  LBL  &  \parbox[t]{2.8cm}{\raggedright Annual Report \vspace*{0.2cm}}  &  \refcite{1991Hen01} \\
\parbox[t]{1.5cm}{\raggedright$^{255}$Db  \vspace*{0.08cm}} &  G. N. Flerov  & 1976 &  Dubna  &  \parbox[t]{2.8cm}{\raggedright  Nuclei Far from Stability 1976 \vspace*{0.08cm}}  &  \refcite{1976Fle01} \\ 
\botrule
\end{tabular}}
\end{table}

As mentioned in the introduction, only discoveries reported in refereed journals were included. Typically, new results are first presented at conferences and brief write-ups are then included in the proceedings, sometimes labeled as preliminary. Quite often preliminary results are also included in internal and/or annual reports. In most instances, the final results are subsequently submitted for publication in regular journals. However, for whatever reasons, in some cases this last step does not occur. At the present time the discovery of 22 nuclides have so far only been reported in conference proceedings or internal reports and are listed in Table \ref{reports}. 
 
From the 26 nuclides listed in the previous review\cite{2013Tho02} eight have been published in refereed journals in 2012 $^{95}$Cd (Ref.~\refcite{2012Hin01}), $^{97}$In (Ref.~\refcite{2012Hin01}), $^{155}$Pr (Ref.~\refcite{2012Van01}), $^{157}$Nd (Ref.~\refcite{2012Van01}), $^{158}$Nd (Ref.~\refcite{2012Kur01}), $^{178}$Tm (Ref.~\refcite{2012Kur01}), and $^{181,182}$Yb (Ref.~\refcite{2012Kur01}). Last year four new nuclides ($^{81,82}$Mo and $^{85,86}$Ru) have been reported in the proceedings of the 2012 EMIS conference\cite{2013Suz01}. In addition, the observation of $^{230}$At and $^{232}$Rn which was reported in a preprint\cite{2010Ben02} and presented at a conference\cite{2010Ben01} in 2010 was included. The only neutron-rich nucleus in the list ($^{156}$Pr) has also been observed and will most likely be published during the next year\cite{2013Kub01} (see next section). 

Of the nine transuranium nuclides on the list only $^{234}$Cm, $^{234}$Bk and maybe $^{235}$Cm 
will potentially still be published in refereed journals. All other internal/annual reports or contributions to conference proceedings are more than twenty years old and most likely will require new experiments to confirm and verify the old results. 

The five nuclei reported by Souliotis ($^{126}$Nd, $^{136}$Gd, $^{138}$Tb,$^{143}$Ho, and $^{150}$Yb) were observed in a fragmentation reaction\cite{2000Sou01}. It is unlikely that these specific results will still be published in a regular journal. However, as already stated in the previous review: ``The recent advances in beam intensities and detection techniques for fragmentation reactions (especially identification and separation of charge states) should make it possible to discover these and many more additional nuclides along and beyond the proton dripline in this mass region.''\cite{2013Tho02}

The two remaining nuclei, $^{144}$Tm and $^{164}$Ir were reported at conferences about 10 years ago, however, in principle there should be no reason why these results cannot still be published in a refereed journal.

\section{Outlook for 2014}
A longer term outlook for the potential of discovering new nuclides has been presented in Ref. \refcite{2013Tho02}. It is not anticipated that the number of new nuclides reported in 2014 will be large. The two new next generation fragmentation facilities - FAIR\cite{2009FAI01} at GSI and FRIB\cite{2010Bol01,2012Wei01} at MSU - will not be ready until at least the end of the decade so that most probably only RIBF\cite{2010Sak01} at RIKEN is in the position to produce a large number of new nuclides within a single experiment. 

Based on presentations at conferences, seminars and private communications it is anticipated that the discovery of about twenty new nuclides will be published in 2014. The first observation of $^{205}$Ac at Lanzhou, China has already been accepted for publication in Phys. Rev. C: ``The new neutron-deficient isotope $^{205}$Ac was synthesized in the complete-fusion reaction $^{169}$Tm($^{40}$Ca, 4n)$^{205}$Ac. The evaporation residues were separated in-flight by the gas-filled recoil separator SHANS in Lanzhou and subsequently identified by the $\alpha$-$\alpha$ position and time correlation method. The $\alpha$-decay energy and half-life of $^{205}$Ac were determined to be 7.935(30) MeV and 20$^{+97}_{-9}$ ms, respectively.''\cite{2014Zha01}

The four nuclides produced in fragmentation reactions at RIBF and reported at EMIS 2012, $^{81,82}$Mo and $^{85,86}$Ru, should be published in a refereed journal this year, however, at the same time the previously reported observation of $^{103}$Sb (Ref.~\refcite{1995Ryk01}) has to be retracted: ``in the measurements with the $^{124}$Xe beam, we have discovered four new isotopes on the proton-drip line, $^{85,86}$Ru and $^{81,82}$Mo, and obtained the clear evidence that $^{103}$Sb is particle unbound with an upper limit of 49~ns for the half-life.''\cite{2013Suz01} In addition, Kubo presented the first observation of the thirteen neutron-rich nuclei $^{153}$Ba, $^{154,155}$La, $^{156,157}$Ce, $^{156-160}$Pr, $^{162}$Nd, $^{164}$Pm, and $^{166}$Sm at the 2013 International Nuclear Physics Conference\cite{2013Kub01} and the results could be ready for publication soon.

In another set of experiments at RIBF, light neutron-rich secondary beams were produced from $^{48}$Ca fragmentation and neutron-unbound nuclei were studied with the SAMURAI/NEBULA\cite{2013Kob01,2012Kon01} setup. First preliminary results indicate resonances in the unbound nuclei $^{21}$C and $^{20}$B produced in one-neutron and one-proton removal reactions, respectively, from a secondary $^{22}$C beam.\cite{2013Mar01} It should be mentioned that a previous search for $^{21}$C from a one-proton knockout reaction from $^{22}$N was unsuccessful \cite{2013Mos01}. There also might be the possibility to reconstruct resonances in the unbound nuclide $^{24}$N from two-proton removal reactions from a secondary $^{26}$F beam\cite{2013Nak01}. 

\section{Summary}

Although no new major facilities came online during the last few of years, a steady number of new nuclides have been reported from several different facilities. This trend is expected to continue for the next few years with an estimated number of about twenty new nuclides per year. Individual new nuclides are still in reach from many facilities around the world as demonstrated by the recent observation at Jyv\"asky\"a\cite{2013Uus01} and Lanzhou,\cite{2014Zha01} and the continuation of the exploration of superheavy nuclei at Dubna.\cite{2013Oga01,2013Oga02} However, in order to continue the positive slope of the discovery rate shown in Figure \ref{f:timeline} it will be necessary to increase the primary beam intensities at the current projectile fragmentation facilities in order to expand the reach towards even more neutron-rich nuclei. Ultimately, with the next generation facilities like FAIR\cite{2009FAI01} and FRIB,\cite{2010Bol01} several hundreds of new neutron-rich nuclides will be discovered.

\section*{Acknowledgements}

I would like to thank Prof. Peter Armbruster for the information regarding $^{235}$Ac and $^{271}$Ds, Dr. Dieter Ackermann and Prof. Christoph E. D\"ullmann for discussions regarding the discovery of $^{266}$Db, Dr. Juan Flegenheimer for correcting the assignment of $^{195}$Os and Ute Thoennessen for carefully proofreading the manuscript. Support of the National Science Foundation under grant No. PHY11-02511 is gratefully acknowledged.

\bibliographystyle{ws-ijmpe}
\bibliography{isotope-references}

\end{document}